\documentclass[aps,pra,twocolumn]{revtex4-1}

\usepackage{graphicx}

\begin{document}

\title{Measurement uncertainties in the quantum formalism:\\
quasi-realities of individual systems
}
\author{Holger F. Hofmann}
\email{hofmann@hiroshima-u.ac.jp}
\affiliation{
Graduate School of Advanced Sciences of Matter, Hiroshima University,
Kagamiyama 1-3-1, Higashi Hiroshima 739-8530, Japan}
\affiliation{JST, CREST, Sanbancho 5, Chiyoda-ku, Tokyo 102-0075, Japan
}

\begin{abstract}
The evaluation of uncertainties in quantum measurements is problematic since the correct value of an observable between state preparation and measurement is experimentally inaccessible. In Ozawa's formulation of uncertainty relations for quantum measurements, the correct value of an observable is represented by the operator of that observable. Here, I consider the implications of this operator-based assignment of values to individual systems and discuss the relation with weak values and weak measurement statistics.
\end{abstract}

\maketitle

\section{Introduction}

In its widely popularized form, the uncertainty principle states that it is impossible to know all properties of a quantum system at the same time. A more careful formulation would be that the quantum formalism does not allow a description of quantum systems in terms of a complete set of physical properties. Whether the unknown properties actually exist or have any physical meaning is a matter of controversy since the early days of quantum mechanics. That this controversy could continue for so long seems to point to a rather peculiar failure of the established scientific methods. From these methods, one would expect that the combination of experiment and theory should eventually result in an unambiguous description of physical reality. However, the uncertainty principle has played a crucial role in preventing a resolution of the controversies: any attempt to uncover more about the physical reality of an individual system than the uncertainty principle allows must fail because of uncontrollable disturbances and errors in the measurement. This problem not only limits the experimental possibilities, but also restricts the statements derived from purely theoretical considerations of possible measurements. Therefore, neither experiment nor theory provides unambiguous statements about physical reality. 

Recent advances in quantum information related technologies have renewed the interest in the fundamental physics described by the quantum formalism. Applications of entanglement seem to indicate that there might be intriguing new ways to circumvent the uncertainty principle. In the light of these developments, it might be possible to resolve the problems associated with uncertainties at their most fundamental level by directly addressing the dynamics of quantum measurements in the Hilbert space formalism. An important contribution to this effort was made by Ozawa, who introduced an uncertainty relation for the error and the disturbance of quantum measurements that includes the initial quantum state of the system \cite{Oza03}. Recently, experiments on sequential spin measurements were used to illustrate the physics described by the general formalism introduced by Ozawa \cite{Erh12}. However, these experiments could not provide direct evidence of the errors and the disturbances, since the actual properties of the system between preparation and initial measurement remain as inaccessible as ever. Instead, the experimentalists used a tomographic reconstruction of the elements contributing to the theoretically predicted error. Whether this analysis is convincing or not depends on ones interpretation of the theory. In the following, I will therefore examine the basic definitions used in Ozawa's theory and identify the implicit assumptions they make about the physics described by the quantum formalism.   

\section{Measurement uncertainties in\\ the Hilbert space formalism}

The most compact formulation of measurement theory is given by a set of measurement operators $\{\hat{M}_m\}$, where each operator represents the effects of a measurement outcome $m$ on an initial state $\mid \psi \rangle$. Specifically, the probability of obtaining the outcome $m$ is given by the squared norm of the Hilbert space vector $\hat{M}_m \mid \psi \rangle$, and the state after the measurement is given by the direction of that vector in Hilbert space. In this theory, the measurement result $m$ can be used to estimate the most likely value of an observable $\hat{A}$ before the measurement by assigning a measurement result $A_m$ to that observable. If the measurement operator $\hat{M}_m \mid \psi \rangle$ commutes with the observable $\hat{A}$, we can establish the correct value of $\hat{A}$ by a final measurement of the outcomes $a$ described by the eigenstates $\mid a \rangle$ of $\hat{A}$. Since the operator $\hat{A}$ describes the assignment of the correct eigenvalue $A_a$ to the eigenstate $\mid a \rangle$, we can derive the average squared error of the result $A_m$ directly from the output states, 
\begin{equation}
\epsilon_{\mbox{\small out}}^2(A) = \sum_m ||(A_m-\hat{A}) \hat{M}_m \mid \psi \rangle||^2.
\end{equation}
This uncertainty is experimentally accessible, since it describes a prediction of $\hat{A}$ for the output state. If $\hat{M}_m \mid \psi \rangle$ and $\hat{A}$ commute, the measurement of $m$ does not change the value of $\hat{A}$, and the final measurement result $A_a$ can be identified with the correct value of $\hat{A}$ before the measurement of $m$. It may therefore be tempting to assume that the correct value of $\hat{A}$ before the measurement of $m$ can always be obtained by simply changing the operator ordering. This is the implicit assumption made by Ozawa in the formulation of the measurement error \cite{Oza03,Erh12}:
\begin{equation}
\epsilon^2(A) = \sum_m || \hat{M}_m (A_m-\hat{A}) \mid \psi \rangle||^2.
\end{equation}
Problematic about this formulation is that the operator $\hat{A}$ cannot be evaluated using a measurement. Therefore, the assignment of the correct value now depends on an interpretation of the Hilbert space formalism that cannot be verified experimentally. 

The same problem arises in the determination of the disturbance of an observable $\hat{B}$ caused by the measurement interaction. A measurement of the actual disturbance requires that the input state is an eigenstate of $\hat{B}$, so that the disturbance can be given by the difference in the eigenvalues between the initial state and a final measurement on the output. In this case, the operator $\hat{B}$ evaluates the initial value of $\hat{B}$ when placed on the right hand side of $\hat{M}_m$, and the final value of $\hat{B}$ when placed on the left hand side of $\hat{M}_m$. Ozawa generalizes this operator assignment of values to define the disturbance as 
\begin{equation}
\eta^2(B) = \sum_m || (\hat{B} \hat{M}_m - \hat{M}_m \hat{B}) \mid \psi \rangle||^2.
\end{equation}
Again, the implicit assumption is that the operator $\hat{B}$ represents the correct value of $\hat{B}$, even when it cannot be evaluated experimentally. 

Significantly, Ozawa's theory of measurement uncertainties depends on the assumption that the operator formalism of Hilbert space gives us access to unobserved physical properties by effectively identifying the precise value of an observable between preparation and measurement. At this point, it may be worthwhile to reflect a bit on the original motivation for the uncertainty principle: that the Hilbert space formalism cannot describe physical objects in terms of their individual physical properties. For this reason, Heisenberg was not actually concerned with the question of whether unobserved values could be reconstructed or estimated based on prior knowledge. In fact, he expressed the opinion that a complete description of the past was still possible. A direct quote to that effect can be found in Heisenbergs introduction of the physical principles of quantum theory (original German with my own translation)\cite{Hei58}: 
\begin{quote}
``Vorher sei jedoch bemerkt, dass die Unbe\-stimmtheitsrelation sich offenbar nicht auf die Vergangenheit bezieht. Denn wenn zunaechst die Elektronengeschwindigkeit be\-kannt ist, dann der Ort genau gemessen wird, so lassen sich auch fuer die Zeit {\it vor} der Ortsmessung die Elektronenorte genau ausrechnen; fuer diese Vergangenheit ist $\Delta q\Delta p$ dann kleiner als der uebliche Grenz\-wert. Diese Kenntnis der Vergangenheit hat jedoch rein spekulativen Charakter, denn sie geht (wegen der Impulsaenderung bei der Ortsmessung) keineswegs als Anfangsbe\-ding\-ung in irgendeine Rechnung ueber die Zu\-kunft des Elektrons ein und tritt ueberhaupt in keinem physikalischen Experiment in Erscheinung.''
\end{quote}
\begin{quote}
``Beforehand it should be noted that the uncertainty relations obviously do not apply to the past. If the velocity of an electron is initially known, and the position is measured with precision after that, then the positions of the electron may be calculated precisely also for the times {\it before} the measurement of position; for this past, $\Delta q \Delta p$ is then smaller than the usual limit. However, this knowledge of the past is merely speculative in nature, since (because of the change of momentum in the position measurement) it can never enter as initial condition into any calculation concerning the future of the electron and does not appear in any physical experiment at all.''
\end{quote}
Here, Heisenberg suggests that the laws of physics may allow us to know the history of an individual particle with precision. However, he continues by dismissing the notion that this might have any effect on experiments, since the implications for future measurement outcomes have been erased by the disturbance of the velocity in the position measurement. As a result, Heisenberg failed to examine whether the Hilbert space formalism might have any concequences for the classical notions of causality that he uses to explain the physics of the measurement process. 

\section{Values for unobserved observables}

As Heisenberg suggested, it may be possible to determine the precise values of unobserved observables from the initial preparation and the final measurement result obtained for each individual system. In Ozawa's theory, this can be done by adding a final projective measurement $\{\mid f \rangle\}$. The probabilities of the outcomes are then given by
\begin{equation}
p(m,f)= ||\langle f \mid \hat{M}_m \mid \psi \rangle||^2.
\end{equation} 
The measurement error can now be expressed as an average over the squared measurement errors for each individual outcome $(m,f)$,
\begin{equation}
\epsilon^2(A) = \sum_{m,f} p(m,f) \bigg|\bigg| A_m - \frac{\langle f \mid \hat{M}_m \hat{A}\mid \psi \rangle}{\langle f \mid \hat{M}_m \mid \psi \rangle}  \bigg|\bigg|^2.
\end{equation}
Ozawa's theory therefore implies that the correct value of $\hat{A}$ between preparation and measurement is given by the {\it complex} weak value of $\hat{A}$ conditioned by both the initial measurement result $m$ and the final measurement result $f$. 

The fact that Ozawa's theory is closely related to weak values was first pointed out by Hall, who noted that the best estimate $A_m$ for a given measurement result $m$ was given by the real part of the weak value defined by the initial state and the measurement of $m$ \cite{Hal04}. Interestingly, we can now see that, for projective measurements, the error of an estimate given by the real part of the weak value is exactly equal to the imaginary part of the weak value \cite{Hof11a}. Effectively, the  definition of measurement errors given by Ozawa implies that the complex weak values provide a precise description of the observables between preparation and measurement, so that any difference between the measurement result and the weak value - even if purely imaginary - corresponds to an error in the estimated value of $\hat{A}$. 

The same logic can be applied to the disturbance. Here, an even more interesting case emerges if we consider an arbitrary final measurement, so that neither the initial nor the final value of $\hat{B}$ is actually known. In this case, Ozawa's disturbance can be derived from the differences between the weak values before and after the measurement,
\begin{eqnarray}
\lefteqn{
\eta^2(B) =} \nonumber \\ && 
\sum_{m,f} p(m,f) \bigg|\bigg| \frac{\langle f \mid \hat{B} \hat{M}_m \mid \psi \rangle}{\langle f \mid \hat{M}_m \mid \psi \rangle} - \frac{\langle f \mid \hat{M}_m \hat{B}\mid \psi \rangle}{\langle f \mid \hat{M}_m \mid \psi \rangle}  \bigg|\bigg|^2.
\end{eqnarray}
Thus, the situation before the measurement is fully characterized by the initial state $\mid \psi \rangle$ and the final state $\langle f \mid \hat{M}_m$, while the situation after the measurement is fully characterized by the initial state $\hat{M}_m \mid \psi \rangle$ and the final state $\langle f \mid$. 

In the light of Heisenbergs considerations, it would seem that quantum theory makes precise and uncertainty free statements about physical reality whenever the initial state and the final state are given by pure states. If Ozawa's definitions of uncertainties are correct (and they are consistent with the conventional use of Operators in Hilbert space), we have to conclude that such precise statements can be represented by the weak values of the operators that describe the physical system in question. However, this raises a problem for Heisenberg's assumption about the physics of a system between preparation and measurement: the quantum formalism assigns complex values to the operators, and even the real parts are not consistent with the eigenvalues observed in strong measurements. It seems that the quantum formalism has much more serious consequences for the description of unobserved observables than Heisenberg anticipated.

\section{Quasi-realities and contextuality}

Experimentally, the problem with weak values is that they cannot be confirmed by precise measurements on individual systems. In fact, weak values appear to contradict the initial justification of the operator formalism, where an operator is seen as a statistical tool that attaches an eigenvalue to the measurement result represented by a projection on the eigenstate. How can it be that the weak values appear as uncertainty free values in a theory based on the standard operator formalism, when they do not correspond to any of the alternative measurement realities represented by the projectors on the eigenstate?

These considerations lead us to an alternative interpretation of Ozawa's uncertainties given by Lund and Wiseman \cite{Lun10}, where the correct values of $\hat{A}$ and $\hat{B}$ are described by the eigenvalues associated with the eigenstate projectors. Since the expectation values of the projectors correspond to the probabilities of the measurement outcomes, their weak values can be identified with conditional probabilities. The quantum state $\mid \psi \rangle$ can then be expressed by a joint probability distribution for the precise measurement outcome $\mid a \rangle$ and the actual measurement outcome $(m,f)$,
\begin{equation}
p(a,m,f) = \langle f \mid \hat{M}_m \mid a \rangle \langle a \mid \psi \rangle \langle \psi \mid \hat{M}^\dagger_m \mid f \rangle.
\end{equation}
As shown by Lund and Wiseman, the measurement error $\epsilon^2(A)$ can then be determined by assigning a (possibly negative) statistical weight equal to the real part of $p(a,m,f)$ to the squared errors given by the differences between the eigenvalues $A_a$ and the measurement outcomes $A_m$,
\begin{equation}
\epsilon^2(A) = \sum_{a,m,f} \mathrm{Re}\left( p(a,m,f) \right) \left(A_m-A_a\right)^2.
\end{equation}
Instead of associating a single correct value with the actual measurement outcome $(m,f)$, this relation introduces a quasi-reality defined by the joint assignment of reality to the measurement outcomesv$(m,f)$ and $\mid a \rangle$. In the case of the disturbance $\eta^2(B)$, the different measurements are given by the eigenstates $\mid b_i \rangle$ for $\hat{B}$ before the measurement, and the eigenstates $\mid b_f \rangle$ for $\hat{B}$ after the measurement. Ozawa's definition of the disturbance can then be expressed in terms of a joint probability for different outcomes $b_i$ and $b_f$ before and after the measurement of $m$,
\begin{equation}
\eta^2(B) = \sum_{b_i,m,b_f} \mathrm{Re}\left( p(b_i,m,b_f) \right) \left(B_f-B_i\right)^2.
\end{equation}
Intuitively, these relations would suggest the simultaneous assignments of well-defined measurement outcomes to non-commuting properties. However, the joint probabilities associated with such an assignment are generally complex, with non-positive real parts. The joint assignment of outcomes to non-commuting measurements therefore corresponds to a quasi-reality that can never be observed in any actual experiment.

\section{Quantum determinism}

It may seem confusing that the identification of weak values with the precise values of an unobserved observable is consistent with the result obtained from a statistical distribution over the eigenvalues of the same observable. To understand how the precision of weak values can be reconciled with the complex valued statistics over eigenvalues, it is necessary to take a closer look the fundamental characteristics of weak measurement statistics \cite{Hos10,Hof11b,Hof12}. For the present discussion, it might be sufficient to consider the conditional measurement error $\epsilon^2(A|m,f)$ derived from the complex conditional probabilities $p(a|m,f)$,
\begin{eqnarray}
\epsilon^2(A|m,f) \!&=&\! \sum_a \mathrm{Re}\left( p(a|m,f) \right) \left(A_m-A_a\right)^2
\nonumber \\ \!&=&\!  \mathrm{Re}\left( \frac{\langle f \mid \hat{M}_m (A_m-\hat{A})^2 \mid \psi \rangle}{\langle f \mid \hat{M}_m \mid \psi \rangle} \right).
\end{eqnarray}
Thus the contribution to the measurement error $\epsilon^2$ for a measurement outcome $(m,f)$ is given by the real part of the weak value of $(A_m-\hat{A})^2$. This is different from the error given by the squared difference between $A_m$ and the weak value of $\hat{A}$, since the weak value of $\hat{A}^2$ is different from the square of the weak value of $\hat{A}$. This difference corresponds to a definition of weak value uncertainties,
\begin{eqnarray}
\lefteqn{\Delta A^2_{\mathrm{weak}}(m,f) =} \nonumber \\ && \mathrm{Re} \left(\sum_a  p(a|m,f) \left(\frac{\langle f \mid \hat{M}_m \hat{A} \mid \psi \rangle}{\langle f \mid \hat{M}_m \mid \psi \rangle} - A_a \right)^2 \right). 
\end{eqnarray} 
Since the average of these weak value uncertainties must be zero for all $m$ and $f$, the individual values are both negative and positive. This is a consequence of the complex valued joint probabilities that describe the actual relations between measurement outcomes that cannot be obtained at the same time. 

It is remarkable that the use of complex probabilities can establish a relation between the seemingly deterministic definition of uncertainty free values and the statistical formalism of conditional probabilities. As pointed out in \cite{Hof11b}, the precise connection is described by the complex phases of the weak conditional probabilities, which are related to the classical distance between the phase space points associated with initial, final, and intermediate states. Quantum mechanics therefore results in a fundamental modification of classical determinism that reconciles the apparent randomness of quantum effects with continuous state evolutions \cite{Hof12}. In the classical limit, we can determine the value between preparation and measurement from initial and final conditions using the known equations of motion (as suggested by Heisenberg for the electron with known initial momentum and final position). In quantum mechanics, the best approximation to classical determinism seems to be given by weak values. However, a more detailed analysis reveals complex valued probabilities that describe the relations between different measurement contexts. As shown in \cite{Hof12}, classical determinism emerges from these complex valued statistics in the limit of low resolution. The extension of statistics to complex valued probabilities can therefore lift the contradictions between deterministic theories and probabilistic theories, resulting in a contextual determinism that is expressed in terms of statistical relations with an average uncertainty of zero. 

\section{Physics and formalism:\\the lack of a microscopic picture}

In classical physics, we can use the known relations between different quantities to calculate the values of unobserved observables. This calculation reflects a belief in the existence of a microscopic measurement independent reality, where the state of a system is described by a phase space point that summarizes the precise values of all physical properties. In quantum mechanics, the operator representation obscures the relation between different physical properties. Instead of phase space points, quantum states represent probability distributions. It would therefore be fair to argue that conventional quantum mechanics does not provide any microscopic picture of individual physical systems. However, as Heisenberg indicated, such a microscopic picture should re-emerge when we discuss the properties of a system between preparation and measurement. 

Ozawa's approach to uncertainties in quantum measurements is based on the assumption that the operator formalism should also describe the unobserved observables of a system. The result corresponds to the assignment of weak values to the state between preparation and measurement. However, this assignment does not provide a microscopic picture of reality. In particular, the assignment of truth values to precise measurement outcomes does not result in a well-defined assignment of ``0'' or ``1'', but needs to be represented by complex conditional probabilities. 

Significantly, the complex conditional probability can describe uncertainties of zero without specifying a particular outcome. In this case, the relation between the initial, the final, and the unobserved property of the system is precisely defined in terms of complex conditional probabilities, such that there cannot be any joint reality of all three. For individual systems, the maximal amount of information is the combination of initial and final conditions represented by a pair of pure states. Incidentally, this corresponds to the definition of a classical phase space point by a pair of conjugate variables \cite{Hof12}. Ozawa's definitions indicate that the uncertainties of all unobserved observables can then be reduced to zero by evaluating their weak values. However, the overlap with eigenstates of any unobserved property is given by a complex conditional probability. This indicates that contextuality is now a fundamental part of the description: the realities of individual systems cannot be represented by a summary of all physical properties, but must include a specific reference to the properties defined by the past and the future of the system in question. 

If the operator formalism is accepted as a valid representation of physical properties, it is also necessary to accept this contextual notion of microscopic reality. Significantly, the formalism does provide a unique definition of microscopic reality for individual systems, once we agree to take the fundamental definitions to their logical conclusions. This may be the most surprising outcome of the present investigation, since the conventional formulation of quantum mechanics is deliberately unclear and ambiguous about the actual relation between non-commuting observables and the necessary modifications of the phase space picture of classical physics associated with them. 

\section{Conclusions}

Ozawa's formulation of uncertainties is interesting because it derives statements about unobserved observables from the standard formulation of quantum mechanics. This derivation suggests that the standard formalism is much more specific about the relation between physical properties than our textbook knowledge would have us believe. Indeed, a closer look at the mathematics reveals that the statistics of weak measurement provides a key to unlocking the mysteries of uncertainties: Ozawa's formulation shows that the standard formalism assigns weak values and weak conditional probabilities to unobserved observables, since this is the only assignment consistent with the representation of measurement realities by projection operators. 

It is important to note that this assignment is precise in the sense that its average uncertainties are zero. It therefore corresponds to classical expectations such as the one expressed by Heisenberg about the derivation of an electron trajectory from position and momentum in \cite{Hei58}. At the same time, quantum mechanics prevents the identification of unobserved realities by representing such precise relations in terms of complex conditional probabilities. It should therefore be possible to formulate a more detailed microscopic explanation of quantum mechanics by showing how such complex conditional probabilities can represent deterministic relations between non-commuting physical properties \cite{Hof12}.

\vspace{0.3cm}

Part of this work was supported by a Grant-in-Aid of the Japan Society for the Promotion of Science, JSPS.

\end{document}